\documentclass[12pt,manuscript]{aastex}
\usepackage{apjfonts}

\newcommand{\gapprox}{\mathrel{\mathpalette\@versim>}}
\newcommand{\lapprox}{\mathrel{\mathpalette\@versim<}}
\newcommand{\propapprox}{\mathrel{\mathpalette\@versim\propto}}

\shorttitle{Density Variations in Tycho} 
\shortauthors{WILLIAMS ET AL.}

\begin{document}

\title{An X-ray and Radio Study of the Varying Expansion Velocities in
  Tycho's Supernova Remnant}

\author{Brian J. Williams,\altaffilmark{1}
Laura Chomiuk,\altaffilmark{2}
John W. Hewitt,\altaffilmark{3}
John M. Blondin,\altaffilmark{4}
Kazimierz J. Borkowski,\altaffilmark{4}
Parviz Ghavamian,\altaffilmark{5}
Robert Petre,\altaffilmark{6}
Stephen P. Reynolds,\altaffilmark{4}
}

\altaffiltext{1}{CRESST/USRA and X-ray Astrophysics Laboratory, NASA GSFC, 8800 Greenbelt Road, Greenbelt, MD, Code 662, brian.j.williams@nasa.gov}
\altaffiltext{2}{Department of Physics and Astronomy, Michigan State University, East Lansing, Michigan 48824, USA}
\altaffiltext{3}{University of North Florida, Department of Physics, 1 UNF Drive, Jacksonville, FL 32224, USA}
\altaffiltext{4}{Department of Physics, North Carolina State University, Raleigh, NC 27695}
\altaffiltext{5}{Department of Physics, Astronomy, and Geosciences, Towson University, Towson, MD 21252}
\altaffiltext{6}{NASA GSFC, X-ray Astrophysics Laboratory, Greenbelt, MD 20771, USA}

\begin{abstract}

We present newly obtained X-ray and radio observations of Tycho's
supernova remnant using {\it Chandra} and the Karl G. Jansky Very
Large Array in 2015 and 2013/14, respectively. When combined with
earlier epoch observations by these instruments, we now have time
baselines for expansion measurements of the remnant of 12-15 year in
the X-rays and 30 year in the radio. The remnant's large angular size
allows for proper motion measurements at many locations around the
periphery of the blast wave. We find, consistent with earlier
measurements, a clear gradient in the expansion velocity of the
remnant, despite its round shape. The proper motions on the western
and southwestern sides of the remnant are about a factor of two higher
than those in the east and northeast. We showed in an earlier work
that this is related to an offset of the explosion site from the
geometric center of the remnant due to a density gradient in the ISM,
and using our refined measurements reported here, we find that this
offset is $\sim 23''$ towards the northeast. An explosion center
offset in such a circular remnant has implications for searches for
progenitor companions in other remnants.


\end{abstract}

\section{Introduction}
\label{intro}

Tycho's supernova remnant (SNR; hereafter Tycho), is the remnant of
the SN observed in 1572 \citep{stephenson02}, first characterized by
\citet{baade45} as a ``Type I'' SN. \citet{rest08} identified light
echoes from the event, and spectroscopy of these echoes by
\citet{krause08} matched the spectrum to a normal SN Ia. Many authors
have adopted a distance of 2.3 kpc, suggested by \citet{chevalier80}
based on an analysis of optical observations of nonradiative shocks,
and by \citet{albinson86}, based on H I absorption. Distances of over
4 kpc have also been reported, based on higher-resolution H I data
\citep{schwarz95}. In an earlier work (\citealt{williams13}, hereafter
W13), we favored a distance of 3.5 kpc, based on comparisons of
hydrodynamic simulations with data. There, we examined the infrared
(IR) colors of the remnant, fitting models of warm dust to the {\it
  Spitzer} broad-band fluxes. These models are sensitive to the
post-shock gas density, and we found a variation in density as a
function of azimuthal angle around the shell, with densities in the
east and northeast several times higher than in the west and
southwest.

SNe Ia are believed to result from a thermonuclear explosion of a
white dwarf destabilized by mass transfer in a binary system, but the
nature of the binary is unknown. The two leading models are the
single-degenerate (SD) and double-degenerate (DD) scenarios. In the SD
channel \citep{whelan73}, a white dwarf accretes matter from a
non-degenerate companion and explodes when it reaches the
Chandrasekhar limit. In this scenario, the companion star should
survive, though evidence for this via searches for this companion star
in Type Ia SNRs is scant \citep{schaefer12}. The DD channel
\citep{webbink84}, on the other hand, involves an explosion triggered
by the merger of two white dwarfs, and no surviving companion is
expected.

While Tycho is clearly round in shape, a departure from spherical
symmetry is suggested by measurements of the proper motion of the
shell. \citet{reynoso97} used radio images from 1983/84 and 1994/95
and found that the shock velocity varies by a factor of three as a
function of position around the shell. \citet{hughes00} confirmed
these azimuthal variations from {\it ROSAT} images. \citet{katsuda10}
examined {\it Chandra} images from 2000, 2003, and 2007 to measure the
expansion of the remnant at 39 positions around the periphery and
determined that the X-ray proper motions of the forward shock vary by
about a factor of two.

The radio and X-ray emission from the forward shock in Tycho both
arise from the same physical process, nonthermal synchrotron
emission. Electrons are accelerated by turbulent magnetic fields,
amplified at the shock front. This acceleration produces a nonthermal
``tail'' to the particle energy distribution, where electrons reach
relativistic energies up to tens of TeV. These highest energy
electrons, spiraling in the magnetic field, produce X-ray synchrotron
emission \citep{reynolds99}, while the radio emission results from
particles with energies in the GeV range. While most young remnants
exhibit thermal X-ray emission at their forward shock, \citet{hwang02}
show that in Tycho, the emission at the immediate edge is nonthermal.

In this work, we report on proper motion measurements of the forward
shock, made from newly-obtained X-ray and radio data from {\it
  Chandra} and the {\it Karl G. Jansky Very Large Array (VLA)} in 2015
and 2013/14, respectively. While the measurements made by
\citet{reynoso97} and \citet{katsuda10} follow the same general trend,
there are discrepancies between them. This is not unexpected, as these
were independent groups of authors using different techniques with
relatively short time baselines. With much longer time baselines
(twice as long in the X-ray band and three times longer in the radio),
as well as a consistent approach to the measurements in both, we can
greatly reduce the uncertainties and disparities in the previous
measurements. Accurate determinations of shock velocities in Tycho are
important for the physics of particle acceleration, as electron
maximum energies and efficiencies of acceleration and magnetic-field
amplification depend on high powers of the velocity.

The X-ray and radio profiles have different shapes, owing to the
physics of synchrotron emission. X-ray synchrotron emission is
confined to a thin rim that rises sharply at the shock front, only to
fall again fairly quickly behind it as a result of synchrotron losses
on the electrons, damping of the magnetic field, or both. This
phenomenon is discussed at length in \citet{ressler14} and
\citet{tran15} (for Tycho in particular). Radio synchrotron emission
results from lower energy electrons with a longer life, and persists
well behind the shock front, creating a plateau emission profile.

\section{Observations}
\label{obs}

\subsection{X-ray}

{\it Chandra} has observed Tycho four times prior to our 2015
observations: 50 ks in 2000, 150 ks in 2003 and 2007, and 750 ks in
2009. The 2000 observation used the ACIS-S3 chip, and the positioning
of Tycho on the chip resulted in about 25\% of it along the southern
shell being cut off by the chip edge \citep{hwang02}. Subsequent
observations used the I-array and cover the entire remnant. For this
work, we use the longest time baselines. For most of the remnant, our
measurements are made between the 2000 and 2015 data. For the portions
in the south that were not covered by the 2000 observation
(approximately one third of our regions), we substitute the 2003
epoch.

\begin{figure}[htb]
\includegraphics[width=16cm]{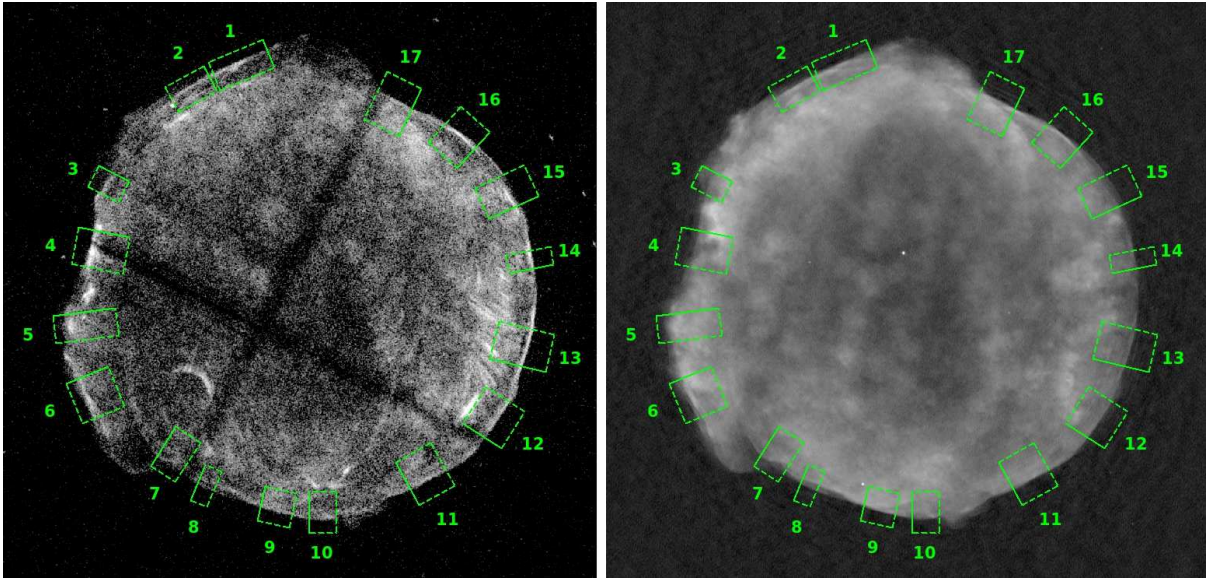}
\caption{Left: {\it Chandra} 3-8 keV X-ray image of Tycho from 2015,
  smoothed with a 2-pixel Gaussian. Right: VLA image from
  2013. Overlaid on both are the 17 regions in which the proper motion
  of the shock is measured. Images are 10.8$'$ on a side.
\label{images}
}
\end{figure}

We observed Tycho for 150 ks on 2015 Apr 22-24 using the ACIS
I-array. The time intervals from the 2000 and 2003 observations to the
2015 are 14.6 and 12.0 year, respectively. We follow a similar data
reduction procedure to \citet{katsuda10}, using version 4.7 of CIAO
and version 4.6.5 of CALDB to process all epochs. We examined the
light curves and found no significant background flares. To align all
epochs to a common reference frame, we first used the CIAO task
\verb|wavdetect| to detect point sources in the field. The tasks
\verb|wcs_match| and \verb|wcs_update| then reprojected each events
file to identical wcs coordinates. We use the deep 2009 observation as
the relative ``reference'' frame to which all other epochs are
aligned. We smooth the X-ray images slightly, using a 2-pixel
Gaussian. This has virtually no effect on the profile shapes, but
significantly decreases the pixel-to-pixel Poisson noise level.

\subsection{Radio} 

We imaged two epochs of L-band ($\sim$1.4 GHz) observations, both
obtained with the VLA in its A, B, C, and D configurations. The first
epoch (PI W. van Breugel), from 1983--1984, is published in
\citet{reynoso97}, but is re-imaged here. The second was obtained by
our team as part of project VLA/13A-426 (PI J.W. Hewitt), on 2013 Feb
9 (D configuration), 2013 Jun 8 (C configuration), 2013 Sep 28 (B
configuration) and 2014 Feb 20 (A configuration). For simplicity, we
will refer to these as the ``2013" observations (and the first epoch
as ``1983"). Two other VLA epochs (1994 and 2002) have also been
obtained, but we desire the longest time baseline, and only use the
1983 and 2013 data.

Data are edited and calibrated using standard routines in AIPS, and
data from all configurations were concatenated in the uv plane using
\verb|VBGLU| for the 1983 epoch and \verb|DBCON| for the 2013
epoch. In both epochs, the source J0217+7349 was used for complex gain
calibration, while 3C48 was used for absolute flux calibration and, in
the case of the 2013 epoch, bandpass calibration. We imaged in AIPS
using multi-scale clean in \verb|IMAGR|.

Data in the A configuration were obtained over 1983 Nov 13--14,
providing 5.4 hours on source with bandwidth of 3.1 MHz at each of two
intermediate frequencies (IFs); IFs were tuned to 1375 and 1385 MHz. B
configuration data were observed 1984 Jan 5--7, yielding 4.6 hours on
source with bandwidth of 6.2 MHz $\times$ 2 IFs, and IFs are
positioned at 1365 and 1442 MHz. C configuration data were obtained on
1983 April 17--18, with one IF at 1375 MHz and 25 MHz of bandwidth;
5.1 hours of on-source time were obtained. D configuration data were
obtained on 1983 Jun 18, with 1.8 hours on source and 25 MHz of
bandwidth tuned to 1375 MHz. The narrow bandwidths in A and B
configurations minimize the effects of bandwidth smearing. We imaged
this epoch using a Briggs Robust value of $-3$ (nearly uniform
weighting), yielding a FWHM of the synthesized beam,
$1.39^{\prime\prime} \times 1.31^{\prime\prime}$, at a PA =
28.2$^{\circ}$.

In each of our 2013 configurations, observations yielded 0.5 hour on
source and made use of the full 1 GHz bandwidth of the L
band. However, to effectively compare with the 1983 epoch, we only
made use of the sixth and seventh spectral windows, which are 64 MHz
wide and centered at 1346 and 1410 MHz. We imaged these data with a
Briggs Robust value of $-5$ (uniform weighting) to produce an image
with resolution $1.86^{\prime\prime} \times 0.88^{\prime\prime}$ at a
PA = 274.7$^{\circ}$. Because the observations from the two epochs had
different point spread functions, and because both of these were
elliptical, we degrade both images to a circular beam of
$1.91^{\prime\prime} \times 1.91^{\prime\prime}$ using the AIPS task
\verb|CONVL|. Both the X-ray and radio images are shown in Figure 1.

\section{Results}
\label{results}

In choosing our measurement regions, we attempt to best match our
previous regions in W13, with some modifications necessitated by the
data, such as choosing locations where the shock front is ``sharpest''
(little diffuse emission). We use X-ray data only in the 3-8 keV range
for two reasons. First, this eliminates virtually all thermal
emission, selecting only the nonthermal synchrotron radiation that
defines the shock front. Secondly, this allows for a more direct
comparison with the radio data, where the emission is also
synchrotron. Our regions are ``projection'' boxes in
ds9\footnotemark[7], and range in width from 20-60$''$, chosen to lie
along the shock front. We made our boxes large enough to provide good
statistics within the region of interest, but small enough so that the
curved shock front remains mostly straight. There are two gaps in our
coverage: one at a PA (E of N) of $\sim 45^{\circ}$ (between boxes 2
and 3) and the other at $\sim 350^{\circ}$ (between boxes 17 and
1). In these locations, the shock front is too diffuse to get a good
profile and a robust proper motion measurement.

\footnotetext[7]{http://ds9.si.edu/site/Home.html}

\begin{figure*}[htb]
\includegraphics[width=8cm]{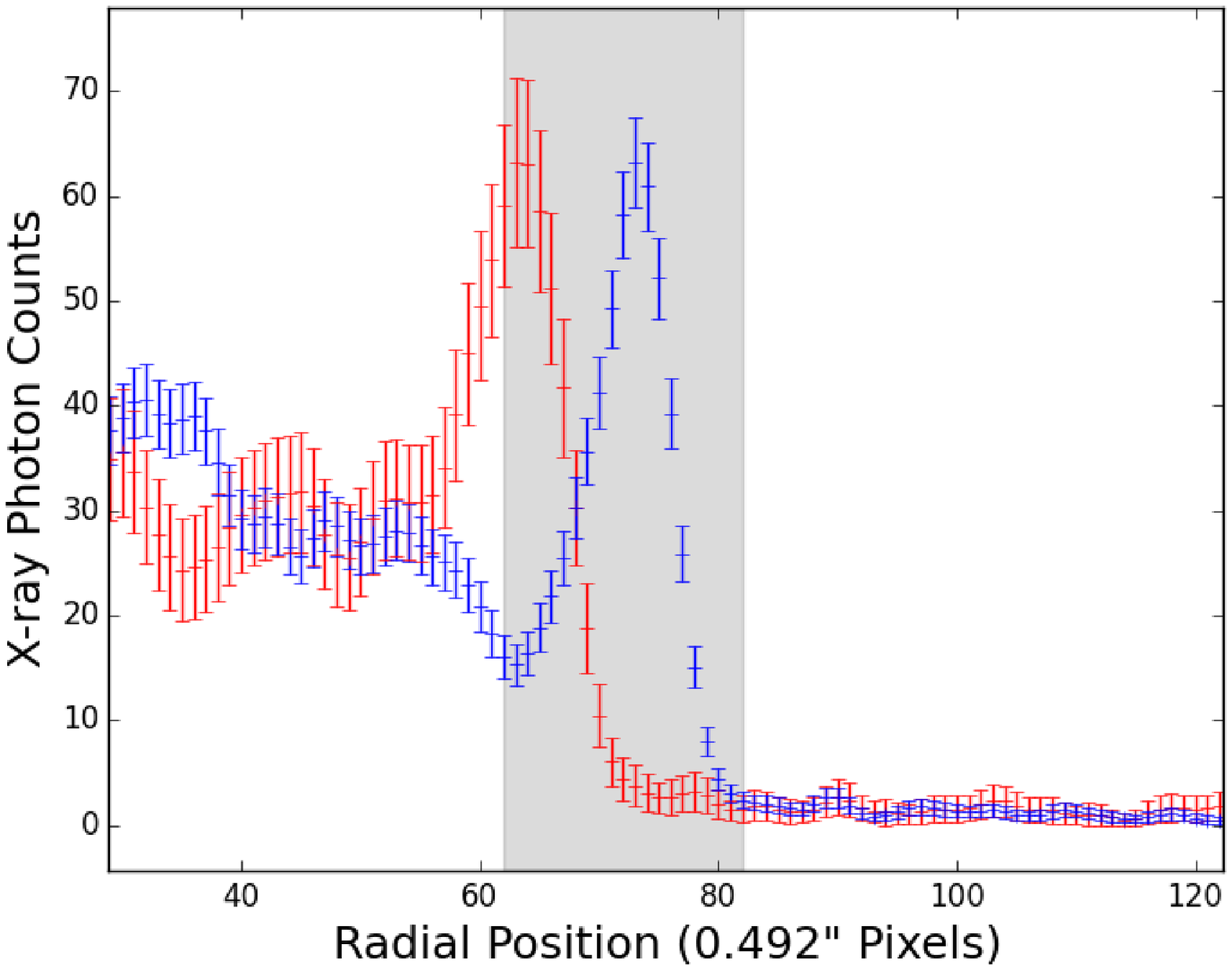}
\includegraphics[width=8cm]{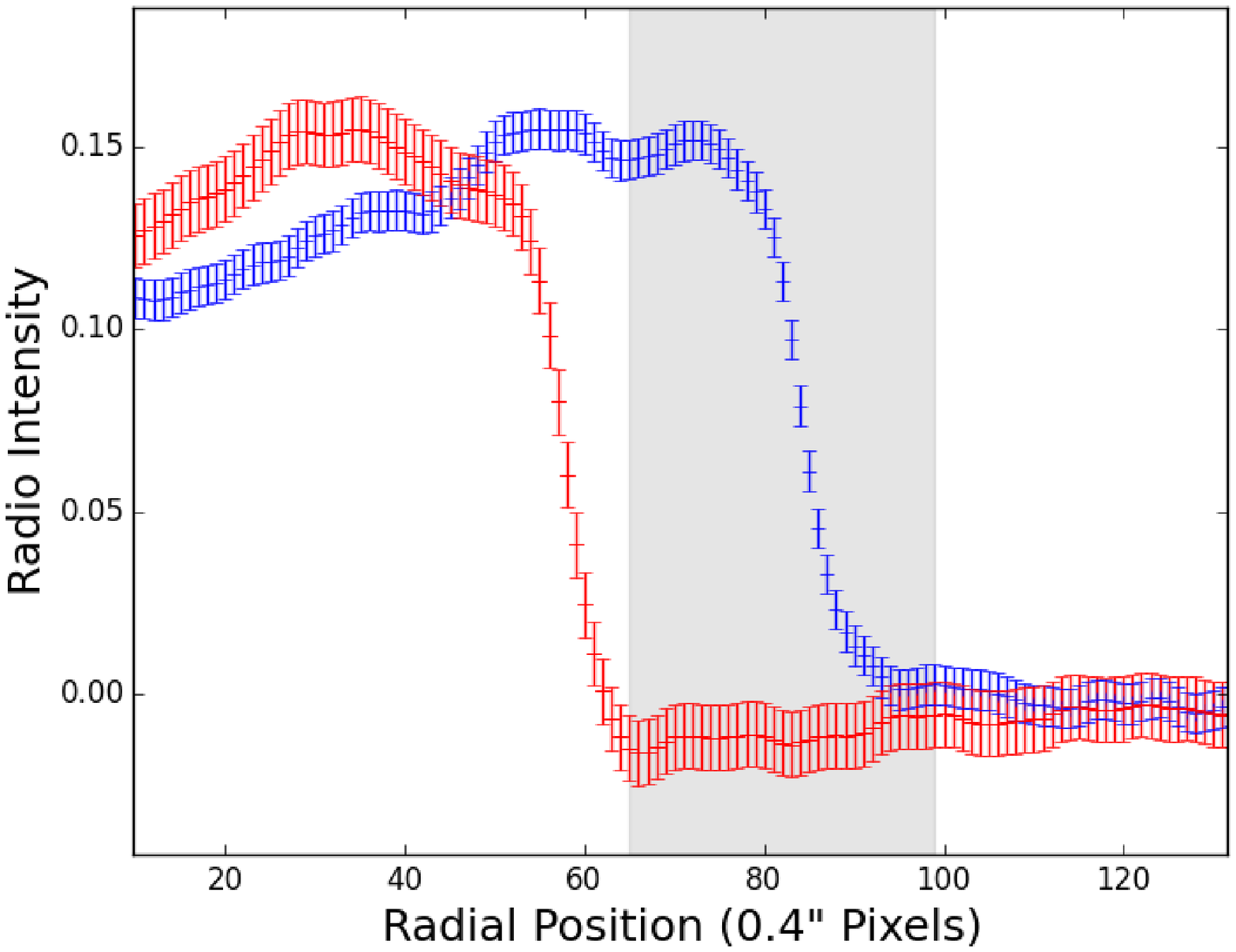}
\caption{Two examples of ``well-behaved'' profiles. In all plots, the
  first epoch is shown in red and the second in blue. {\it Left}: An
  example of an X-ray profile (region 13; P.A. 255$^{\circ}$). {\it
    Right}: An example of a radio profile (region 11,
  P.A. 213$^{\circ}$). Shaded areas mark regions where the fit was
  performed.
\label{goodprofile}
}
\end{figure*}

Our procedure for measuring the proper motion mirrors that found in
\citet{katsuda08a}, where a fuller description can be found. Other SNR
works, such as \citet{winkler14} and \citet{yamaguchi16}, have used
this technique as well. We extract the 1D radial profiles from both
epochs, with uncertainties on each data point, then shift epoch 1
relative to epoch 2, minimizing the value of $\chi^{2}$. Profiles are
extracted in pixel space (X-ray pixel: 0.492$''$, radio pixel:
0.4$''$), with shifts calculated on a grid of 2000 points with a size
of 0.025 pixels. We only fit for the shift within an area (specific to
each region) containing the filament edge, as shown in
Figure~\ref{goodprofile}.

While we use identical regions in both the X-ray and radio, there are
slight differences in the way that we calculate the uncertainties on
the profiles. For the X-ray band, we work in photon counts, and take
the square root of the number of counts in each pixel as the
uncertainty. The radio images are in units of Jy beam$^{-1}$, and we
consider the ``noisiness'' of the image to be the main source of
uncertainty. We calculate the rms dispersion within the off-source
background, effectively a flat-field. We obtain uncertainties of 8.5
$\times 10^{-5}$ and 4.0 $\times 10^{-5}$ Jy beam$^{-1}$ for the 1984
and 2013 images, respectively, and apply these constant uncertainty
values to all points along their respective radio profiles. For both
the X-ray and radio profiles, reduced $\chi^{2}$ values were generally
close to 1.

For the uncertainties on the proper motion measurements themselves, we
include estimates of both the statistical and systematic
uncertainties. The statistical errors are the 90\% confidence limits
resulting from a $\chi^{2}$ increase of 2.706. The 90\% uncertainties
in each direction are virtually identical, so we average them for a
single number. For the systematic errors, the uncertainties arising
from different WCS alignments are negligible. However, what is not
trivial is the angle of the projection box to the shock
front. Following a procedure similar to \citet{katsuda08b}, we vary
the angle of the projection box by 1 and 2$^{\circ}$ in both
directions with respect to the shock front for each region and
re-measure the proper motion. To be conservative, we adopt the
2$^{\circ}$ variation as the measure of our systematic uncertainty. As
with the statistical uncertainties, because the values in each
direction are similar, we average them together for a single value.

\begin{figure*}[htb]
\includegraphics[width=8cm]{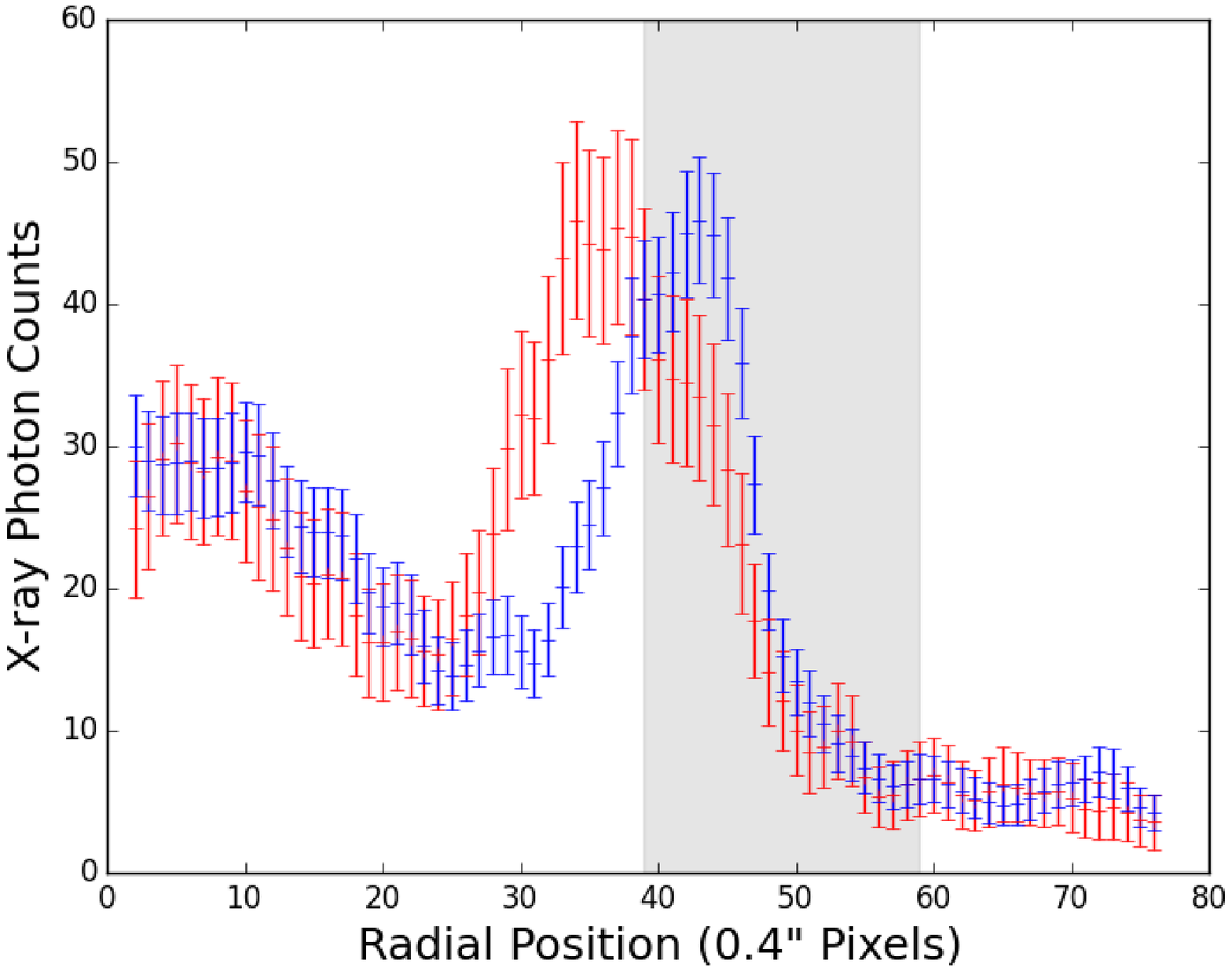}
\includegraphics[width=8cm]{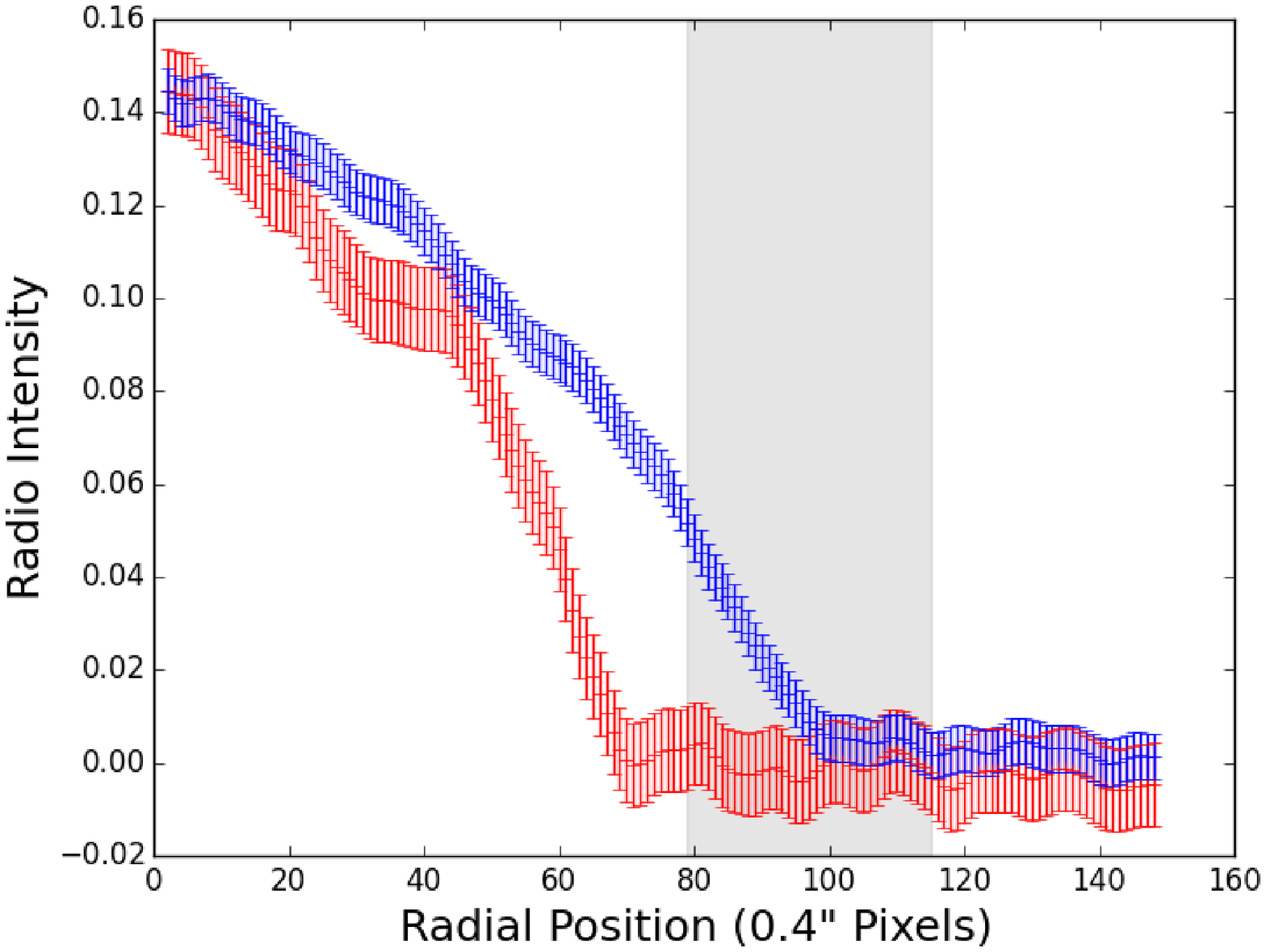}
\caption{The worst-case profiles in X-ray and radio. {\it Left}: The
  X-ray profiles from region 3 (P.A. 62$^{\circ}$). {\it Right}: The
  radio profile from region 15 (P.A. 295$^{\circ}$).
\label{badprofile}
}
\end{figure*}

The vast majority of our regions are ``well-behaved,'' in that the
shape of the profile is virtually identical in both epochs. We show an
example of a well-behaved profile in X-ray and radio in
Figure~\ref{goodprofile}. However, there are a few regions where this
is not the case. We show the worst example from each wavelength regime
in Figure~\ref{badprofile}. In the X-ray band, region 3
(P.A. 62$^{\circ}$) is shown, where the profile clearly changes shape
between the 2000 and 2015 data. We do report a proper motion here,
defining it as the shift in the leading edge of the shock front, when
the profile rises above the level of the background. This region is
even worse in the radio (not shown), where no clear shock front is
present and the profile simply gently fades into the background. We do
not report a radio proper motion for this region.

Region 15 (PA 295$^{\circ}$) is shown as our ``worst-case'' (with the
exception of region 3, discussed above) radio profile. The profile
becomes more elongated between the two epochs, for reasons that are
unknown. Nonetheless, we apply the same technique as above by fitting
the shift of the leading edge of the shock. This leads to a value of
the proper motion, as reported in Table~\ref{measurements}, but this
value should be used with caution. Interestingly, the X-ray profiles
from this region do not show this behavior.

\begin{figure}[htb]
\includegraphics[width=15cm]{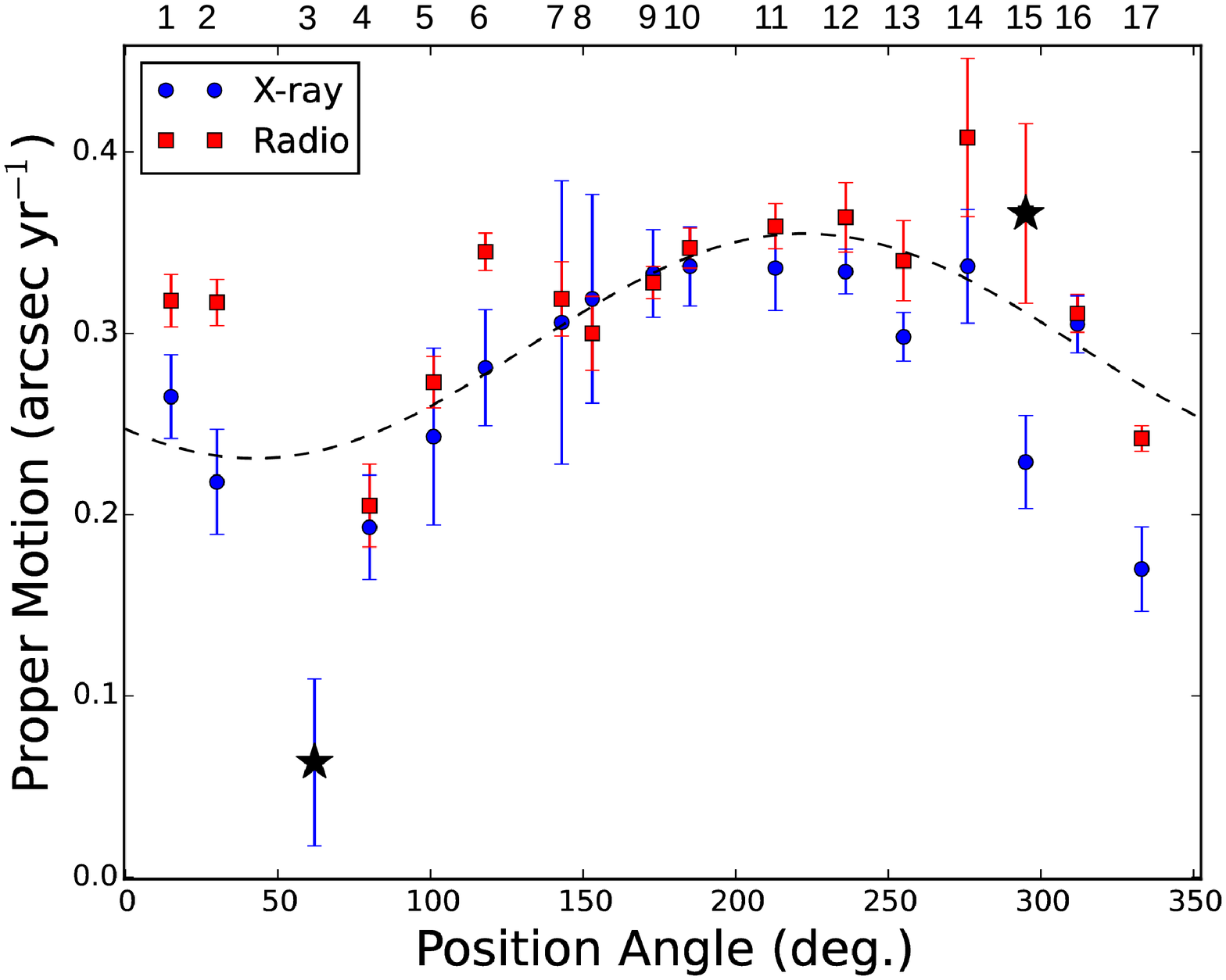}
\caption{The proper motions in all of our regions for both X-rays and
  radio. Region number is given along the top of the plot. Error bars
  plotted are the linear sum of both the statistical and systematic
  uncertainties. The sinusoidal fit described in the text is shown as
  dashed line. Black stars mark the ``bad'' data points shown in
  Figure~\ref{badprofile}.
\label{all}
}
\end{figure}

The results of our measurements are given in Table~\ref{measurements}
and Figure~\ref{all}. We confirm the existence of a clear velocity
gradient in the shock speeds. Even if we disregard the potentially
problematic X-ray measurement of region 3 (discussed above), we still
see a difference of approximately a factor of 2 in the expansion
velocities between the E/NE portions of the remnant and the
W/SW. Furthermore, with a few exceptions, the X-ray and radio proper
motions agree within errors. The biggest discrepancy occurs in region
15, which we have already pointed out is problematic.

While we report all proper motions in units of arcsec yr$^{-1}$, a
conversion to velocity relies upon the distance to Tycho, which may
lie from 2.3 to 4 kpc. If we adopt a fiducial distance of 3 kpc, the
conversion factor is 0.1$''$ yr$^{-1}$ $\equiv$ 1422 km s$^{-1}$. This
distance would imply maximum shock velocities of about 5300 km
s$^{-1}$, consistent with the low pre-shock densities reported in W13,
\citet{katsuda10}, and \citet{cassamchenai07}. It is interesting to
note that our region 4 corresponds to the well-known optical
``knot-G'' of \citet{kamper78}, and the proper motion we measure is
fully consistent with their value of 0.20 $\pm 0.01''$ yr$^{-1}$.

\section{An Off-center Explosion Site}
\label{offcenter}

In W13, we described hydrodynamical simulations of a
spherically-symmetric SN explosion into a density gradient in the ISM,
which is the simplest explanation for the different densities and
shock velocities found around the remnant's periphery. This leads to a
remnant where the geometric center {\it cannot} be the site of the
original explosion. We showed that the remnant can still be remarkably
round, despite higher expansion velocities on one side.

An important result of these simulations, discussed in W13, is that
the ratio of the velocity semi-amplitude (V$_{max}$ -
V$_{min}$)/(V$_{max}$ + V$_{min}$) to the radial offset from the
center of the explosion (R$_{max}$ - R$_{min}$)/(R$_{max}$ +
R$_{min}$) is roughly constant at a value of 2.2 $\pm 0.1$ for ages
between about 300 and 700 year. With our refined proper motion
measurements presented here, we can more accurately calculate the
radial offset necessary to explain the observations.

We use the average of the X-ray and radio proper motions given in
Table~\ref{measurements}, omitting region 3, as this has an unusually
low X-ray proper motion and no measureable radio value. We use a
$\chi^{2}$ minimization algorithm to fit a sinusoidal function, $F =
A\ sin({\theta} + \phi) + Y$ to the data shown in Figure~\ref{all},
for both the amplitude, $A$, and the vertical offset, $Y$, in units of
arcsec yr$^{-1}$, as well as for the phase, $\phi$. In this formalism,
$V_{max}\ = Y+A$ and $V_{min}\ = Y-A$, and the velocity semi-amplitude
reduces to $A/Y$. We obtain best-fit values of A = 0.062 $\pm$ 0.0045
and Y = 0.293 $\pm$ 0.002, leading to a velocity semi-amplitude of
0.212 $\pm$ 0.015. Dividing this by the ratio of 2.2 gives an offset
of (9.6 $\pm\ 0.7$)\% of the radius of the remnant, or (23 $\pm
1.7$)$''$, with the offset in the $V_{min}$ direction, which occurs in
the NE at a position angle of 51$^{\circ} \pm 4.8^{\circ}$ with
respect to the geometric center of \citet{ruiz04}. This is comparable
to the search radius of \citet{kerzendorf13} (who found no viable
candidate for a remaining donor star under the single-degenerate Type
Ia SN scenario), and that of \citet{ruiz04}, who report a possible
candidate. This candidate, ``Star G,'' is 31.4$''$ from our explosion
center.

This is in contrast to \citet{xue15}, who find an offset in the NW
direction. This results in a large ($48.2''$) displacement between our
best-fit explosion center, located at $\alpha$ =
0$^{h}$25$^{m}$22.6$^{s}$ and $\delta$ = 64$^{\circ}$8$'$32.7$''$, and
their results. We note, though, that their paper uses the previously
reported X-ray and radio proper motion measurements of R97 and K10.
Also, our results are based on 2D hydrodynamical simulations instead
of a thin-shell approximation that is strictly valid only for
spherically-symmetric remnants. Whatever the status of the companion
star in Tycho, we stress that a circular morphology does not guarantee
an explosion site in the center of the remnant.

\section{Conclusions}

New observations of Tycho with {\it Chandra} in 2015 and the VLA in
2013/14 have stretched the baselines for proper motion measurements of
the forward shock to 12-15 year and $\sim 30$ year, respectively. We
applied a self-consistent approach to both data sets, refining
previous proper motion measurements. The shock velocity in Tycho
varies by approximately a factor of two from one side of the remnant
to the other, consistent with previous measurements, and with density
variations inferred from {\it Spitzer} observations. This leads to an
offset of about 10\% of the remnant's radius between the geometric
center of the remnant and the site of the explosion. Despite the
circular appearance of Tycho, offsets such as this could exist in
remnants of other SNe as well, impacting the search for surviving
companion stars.

\acknowledgements

Support for this work was provided through Chandra Award GO4-15074Z
issued by the Chandra X-ray Observatory Center, which is operated by
the Smithsonian Astrophysical Observatory for and on behalf of NASA
under contract NAS8-03060. NRAO is a facility of the NSF operated
under cooperative agreement by Associated Universities, Inc. LC
acknowledges NSF AST-1412980.

\begin{deluxetable}{lcccccccc}
\tablecolumns{7} 
\tablewidth{0pc} 
\tabletypesize{\footnotesize}
\tablecaption{Proper Motions as a Function of Position Angle} \tablehead{ \colhead{Reg} & Deg & Width & X-ray PM & $\chi^{2}$ (d.o.f.) & Stat, Syst & Radio PM & $\chi^{2}$ (d.o.f.) & Stat, Syst}

\startdata

1 & 15 & 64 & 0.265 & 7.48 (9) & 0.0104, 0.0127 & 0.318 & 13.9 (13) & 4.59 $\times 10^{-3}$, 9.85 $\times 10^{-3}$\\
2 & 30 & 48 & 0.218 & 8.64 (13) & 0.0224, 6.54 $\times 10^{-3}$ & 0.317 & 16.5 (18) & 4.52 $\times 10^{-3}$, 8.25 $\times 10^{-3}$\\
3 & 62 & 26 & 0.063 & 36.2 (18) & 0.0371, 8.87 $\times 10^{-3}$ & \ldots & \ldots & \ldots \\
4 & 80 & 40 & 0.193 & 12.9 (12) & 0.0104, 0.0184 & 0.205 & 20.7 (22) & 8.33 $\times 10^{-3}$, 0.0145\\
5 & 101 & 30 & 0.243 & 21.3 (28) & 0.0300, 0.0187 & 0.273 & 41.1 (38) & 5.70 $\times 10^{-3}$, 8.46 $\times 10^{-3}$\\
6 & 118 & 47 & 0.281 & 14.8 (15) & 0.0194, 0.0126 & 0.345 & 44.3 (38) & 5.38 $\times 10^{-3}$, 4.84 $\times 10^{-3}$\\
7 & 143 & 32 & 0.306 & 17.8 (22) & 0.0461, 0.0321 & 0.319 & 33.9 (36) & 0.0106, 9.90 $\times 10^{-3}$\\
8 & 153 & 20 & 0.319 & 24.3 (30) & 0.0463, 0.0112 & 0.300 & 26.9 (28) & 6.56 $\times 10^{-3}$, 0.0138\\
9 & 173 & 35 & 0.333 & 15.7 (20) & 0.0168, 7.33 $\times 10^{-3}$ & 0.328 & 36.8 (38) & 4.26 $\times 10^{-3}$, 4.59 $\times 10^{-3}$\\
10 & 185 & 30 & 0.337 & 16.1 (18) & 0.0187, 3.07 $\times 10^{-3}$ & 0.347 & 16.7 (20) & 6.23 $\times 10^{-3}$, 4.85 $\times 10^{-3}$\\
11 & 213 & 44 & 0.336 & 18.3 (23) & 0.0137, 9.74 $\times 10^{-3}$ & 0.359 & 39.7 (33) & 4.92 $\times 10^{-3}$, 7.54 $\times 10^{-3}$\\
12 & 236 & 47 & 0.334 & 20.2 (25) & 9.84 $\times 10^{-3}$, 2.47 $\times 10^{-3}$ & 0.364 & 27.2 (24) & 7.08 $\times 10^{-3}$, 0.0120\\
13 & 255 & 42 & 0.298 & 20.6 (18) & 8.26 $\times 10^{-3}$, 5.07 $\times 10^{-3}$ & 0.340 & 19.0 (22) & 9.90 $\times 10^{-3}$, 0.0122\\
14 & 276 & 20 & 0.337 & 12.2 (16) & 9.77 $\times 10^{-3}$, 0.0216 & 0.408 & 21.0 (18) & 0.0130, 0.0306\\
15 & 295 & 37 & 0.229 & 17.2 (19) & 0.0133, 0.0124 & 0.366 & 121.2 (38) & 0.0220, 0.0275\\
16 & 312 & 42 & 0.305 & 28.7 (24) & 7.63 $\times 10^{-3}$, 4.34 $\pm 0.67$ & 0.311 & 23.6 (20) & 3.93 $\times 10^{-3}$, 6.52 $\times 10^{-3}$\\
17 & 333 & 42 & 0.170 & 16.2 (12) & 0.0147, 8.48 $\times 10^{-3}$ & 0.242 & 17.7 (22) & 3.61 $\times 10^{-3}$, 3.39 $\times 10^{-3}$\\

\enddata

\tablecomments{Deg = Position angle, east of north, with respect to
  geometric center of \citet{ruiz04}. Width of regions in
  arcseconds. Proper motions (PM) in arcseconds yr$^{-1}$. Stat and
  Syst refer to the statistical and systematic uncertainties,
  respectively, on the proper motions.}
\label{measurements}
\end{deluxetable}


\begin{thebibliography}{}

\bibitem[Albinson et al.(1986)]{albinson86}
Albinson, J.S., Tuffs, R.J., Swinbank, E., \& Gull, S.F. 1986, MNRAS, 219, 427

\bibitem[Baade(1945)]{baade45}
Baade, W. 1945, ApJ, 102, 309

\bibitem[Cassam-Chena\"{i} et al.(2007)]{cassamchenai07}
Cassam-Chena\"{i}, G., Hughes, J.P., Ballet, J., \& Decourchelle, A. 2007, ApJ, 665, 315

\bibitem[Chevalier et al.(1980)]{chevalier80}
Chevalier, R.A., Kirshner, R.P., \& Raymond, J.C. 1980, ApJ, 235, 186

\bibitem[Hughes(2000)]{hughes00}
Hughes, J.P. 2000, ApJ, 545, 53

\bibitem[Hwang et al.(2002)]{hwang02}
Hwang, U., Decourchelle, A., Holt, S., Petre, R. 2002, ApJ, 581, 1101

\bibitem[Katsuda et al.(2008a)]{katsuda08a}
Katsuda, S., Tsunemi, H., \& Mori, K. 2008a, ApJ, 678, 35

\bibitem[Katsuda et al.(2008b)]{katsuda08b}
Katsuda, S., Tsunemi, H., Uchida, H., \& Kimura, M. 2008b, ApJ, 689, 225

\bibitem[Katsuda et al.(2010)]{katsuda10} Katsuda, S., Petre, R.,
  Hughes, J.P., Hwang, U., Yamaguchi, H., Hayato, A., Mori, K.,
  Tsunemi, H. 2010, ApJ, 709, 1387

\bibitem[Kamper \& van den Bergh(1978)]{kamper78}
Kamper, K.W. \& van den Bergh, S. 1978, ApJ, 224, 851

\bibitem[Kerzendorf et al.(2013)]{kerzendorf13}
Kerzendorf, W.E., et al. 2013, ApJ, 774, 99

\bibitem[Krause et al.(2008)]{krause08}
Krause, O., Tanaka, M., Usuda, T., Hattori, T., Goto, M., Birkmann, S., \& Nomoto, K. 2008, {\it Nature}, 456, 617

\bibitem[Ressler et al.(2014)]{ressler14}
Ressler, S.M., Katsuda, S., Reynolds, S.P., Long, K.S., Petre, R., Williams, B.J., \& Winkler, P.F. 2014, ApJ, 790, 85

\bibitem[Rest et al.(2008)]{rest08}
Rest, A., et al. 2008, ApJ, 681, 81

\bibitem[Reynolds \& Keohane(1999)]{reynolds99}
Reynolds, S.P. \& Keohane, J.W. 1999, ApJ, 525, 368

\bibitem[Reynoso et al.(1997)]{reynoso97}
Reynoso, E.M., Moffett, D.A., Goss, W.M., Dubner, G.M., Dickel, J.R., Reynolds, S.P., \& Giacani, E.B. 1997, ApJ, 491, 816

\bibitem[Ruiz-Lapuente et al.(2004)]{ruiz04}
Ruiz-Lapuente, P., et al. 2004, Nature, 431, 1069

\bibitem[Schaefer \& Pagnotta(2012)]{schaefer12}
Schaefer, B.E. \& Pagnotta, A. 2012, Nature, 481, 164

\bibitem[Schwarz et al.(1995)]{schwarz95}
Schwarz, U.J., Goss, W.M., Kalberla, P.M., \& Benaglia, P. 1995, A\&A, 299, 193

\bibitem[Stephenson \& Green(2002)]{stephenson02}
Stephenson, F.R. \& Green, D.A. 2002, {\it Historical Supernovae and their Remnants}, Oxford University Press

\bibitem[Tran et al.(2015)]{tran15}
Tran, A., Williams, B.J., Petre, R., Ressler, S.M., \& Reynolds, S.P. 2015, ApJ, 812, 101

\bibitem[Webbink(1984)]{webbink84}
Webbink, R.F. 1984, ApJ, 277, 355

\bibitem[Whelan \& Iben(1973)]{whelan73}
Whelan, J., \& Iben, I., Jr. 1973, ApJ, 186, 1007

\bibitem[Williams et al.(2013)]{williams13}
Williams, B.J., et al. 2013, ApJ, 770, 129

\bibitem[Winkler et al.(2014)]{winkler14}
Winkler, P.F., et al. 2014, ApJ, 781, 65

\bibitem[Xue \& Schaefer(2015)]{xue15}
Xue, Z. \& Schaefer, B.E. 2015, ApJ, 809, 183

\bibitem[Yamaguchi et al.(2016)]{yamaguchi16}
Yamaguchi, H., Katsuda, S., \& Castro, D. et al. 2016, ApJ, 820, 3

\end{thebibliography}
\end{document}